\documentclass[aps,pra,twocolumn,showpacs,superscriptaddress]{revtex4-1}
\usepackage{graphicx}
\usepackage{amsmath}
\usepackage[dvipdfm,colorlinks=true, citecolor=blue, urlcolor=blue, linkcolor=blue ]{hyperref}

\begin{document}
\title{Floquet control of quantum dissipation in spin chains}
\author{Chong Chen}
\affiliation{Center for Interdisciplinary Studies $\&$ Key Laboratory for Magnetism and Magnetic Materials of the MoE, Lanzhou University, Lanzhou 730000, China}
\author{Jun-Hong An}\email{anjhong@lzu.edu.cn}
\affiliation{Center for Interdisciplinary Studies $\&$ Key Laboratory for Magnetism and Magnetic Materials of the MoE, Lanzhou University, Lanzhou 730000, China}
\affiliation{Beijing Computational Science Research Center, Beijing 100084, China}
\affiliation{Centre for Quantum Technologies, National University of Singapore, 3 Science Drive 2, Singapore 117543, Singapore}
\author{Hong-Gang Luo}
\affiliation{Center for Interdisciplinary Studies $\&$ Key Laboratory for Magnetism and Magnetic Materials of the MoE, Lanzhou University, Lanzhou 730000, China}
\affiliation{Beijing Computational Science Research Center, Beijing 100084, China}
\author{C. P. Sun}
\affiliation{Beijing Computational Science Research Center, Beijing 100084, China}
\author{C. H. Oh}
\affiliation{Centre for Quantum Technologies, National University of Singapore, 3 Science Drive 2, Singapore 117543, Singapore}

\begin{abstract}
  Controlling the decoherence induced by the interaction of quantum system with its environment is a fundamental challenge in quantum technology. Utilizing Floquet theory, we explore the constructive role of temporal periodic driving in suppressing decoherence of a spin-1/2 particle coupled to a spin bath. It is revealed that, accompanying the formation of a Floquet bound state in the quasienergy spectrum of the whole system including the system and its environment, the dissipation of the spin system can be inhibited and the system tends to coherently synchronize with the driving. It can be seen as an analog to the decoherence suppression induced by the structured environment in spatially periodic photonic crystal setting. Comparing with other decoherence control schemes, our protocol is robust against the fluctuation of control parameters and easy to realize in practice. It suggests a promising perspective of periodic driving in decoherence control.
\end{abstract}
\maketitle
\section{Introduction}
As a ubiquitous phenomenon in microscopic world, decoherence is a main obstacle to the realization of any applications of quantum coherence, e.g., quantum information processing \cite{Nielsen2010}, quantum metrology \cite{Chin2012}, and quantum simulation \cite{Britton2012}. Many methods, such as feedback control \cite{Feedback, Feedbackexp}, decoherence-free subspace encoding \cite{DFS,DFSexp}, and dynamical decoupling \cite{Viola1998, Viola1999, Du2009, *Wang2011, Lange2010}, have been proposed to beat this unwanted effect. The dynamical decoupling scheme can be generally described by the so-called spectral filtering theory in the first-order Magnus expansion \cite{KK2001, Uhrig2007, Sarma2008, Biercuk2011, Green2013}, which is valid when the control pulses are sufficiently rapid. Based on the spin echo technique, this scheme is widely exploited to suppress dephasing \cite{Viola1998, Du2009, *Wang2011, Lange2010, Uhrig2007}, where the system has no energy exchange with the environment, and classical noises \cite{Sarma2008, Biercuk2011, Green2013}. It requires a high controllability to the system due to its sensitivity to the time instants at which the inverse pulses are applied \cite{Alexander2012}. Furthermore, when dissipation and quantum noises are involved, it generally cannot perform well. Although the dissipation control was partially touched in the original form of the spectral filtering theory \cite{KK2001}, whether the first-Markovian approximation used there can capture well the physics for such time-dependent systems is still an open question. This is because that new time scales would be introduced to the systems by the time-dependent control field, which might invalidate the application of the Markovian approximation.


As a main inspiration, we notice that the decoherence of dissipative systems connects tightly with the energy-spectrum characters of the total system consisting of the system and its environment \cite{Miyamoto2005, Tong2010, Zhang2013}. If a bound state residing in the energy bandgap of the whole system is formed by changing the environmental spectral density, the decoherence of the system can be suppressed. Thus one can artificially engineer the bound state to suppress decoherence of quantum emitters by introducing spatial periodic confinement in a photonic crystal setting \cite{Fujita2005, Englund2005, Jorgensen2011, Leistikow2011}. Here the spatial periodic confinement dramatically alters the dispersion relation of the radiation field of the quantum emitter such that a certain bandgap structure is present in the spectral density. If the frequency of the quantum emitter resides in the bandgap region, then a bound state is formed and thus the decoherence of the emitter can be suppressed. However, in practical solid-state systems, one generally faces that it is hard to manipulate the spectral density via changing the spatial confinement once the material of system is fabricated. Thus a more efficient way in engineering the bound state than changing the spectral density is desired.

Recently, temporal periodic driving has become a highly controllable and versatile tool in quantum control. Many efforts have been devoted to explore non-trivial effects induced by periodic driving on physical systems. It has been proven to play profound role not only in controlling single-quantum-state of microscopic systems \cite{Grossmann1991, Grifoni1998, Luo2014, Lignier2007, *Eckardt2009, Zhou2009, Das2010, Zhou2010, Tong2014} and implementing geometric phase gates in quantum computation \cite{Bermudez2012, Lemmer2013}, but also in generating novel states of matter absent in the original static system \cite{Vorberg2013, Hauke2012, Lindner2011, Rechtsman2013, Cayssol2013, Nakagawa2014, Tong2013, Guo2013}. Different from static systems, periodically driven systems have no stationary states because the energy is not conserved. Due to Floquet theory, they have well-defined quasi-stationary-state properties described by the Floquet eigen-values, which are called quasienergies. The distinguished role of periodic driving in these diverse systems is that the versatility of driving schemes can induce more colorful quasi-stationary-state behaviors than the static case by controlling the quasi-energy spectrum.



In this paper, we explore the possibility of periodic driving on engineering the bound state of a spin-1/2 system interacting with a XX-type coupled spin bath. Via manipulating the quasi-energy spectrum by periodic driving, we find that a Floquet eigenstate with discrete quasienergy, which we name a Floquet bound state (FBS), can be formed within the bandgap of the quasienergy spectrum. We further reveal that the presence of the FBS would dynamically cause the dissipation of the system spin inhibited. The result suggests that we can manipulate the periodicity in a temporal domain instead of the one in a spatial domain to suppress decoherence, which relaxes greatly the experimental difficulty in fabricating periodic confinement in a photonic crystal.

Our paper is organized as follows. In Sec. \ref{model}, we present our model of a periodically driven spin-$1/2$ particle coupled to a spin chain bath and its exact decoherence dynamics. In Sec. \ref{results}, the Floquet theory is used to obtain the quasi-energy spectrum of the whole system. In Sec. \ref{mechanism}, the mechanism of decoherence inhabitation induced by the periodic driving is revealed. The comparisons of this mechanism with the previous methods are also shown in this section. Finally, a summary is given in Sec. \ref{conclusion}.

\section{Model and dynamics}\label{model}
We consider a periodically driven spin-$1/2$ particle interacting with a one-dimensional spin chain, which is composed of $L$ spin-$1/2$ particles coupled via XX-type interactions. The Hamiltonian of the total system is $\hat{H}(t)=\hat{H}_\text{S}(t)+\hat{H}_\text{I}+\hat{H}_\text{E}$ with
\begin{eqnarray}
  \hat{H}_\text{S}(t)&=&{1\over 2}[\lambda+A(t)]  \hat{\sigma}_0^z,~~\hat{H}_\text{I}=\frac{g}{2} (\hat{\sigma}_0^x \hat{\sigma}^x_1 +\hat{\sigma}_0^y \hat{\sigma}^y_1 ),\label{SH}\\
  \hat{H}_\text{E}&=&{\lambda\over 2}\sum_{j=1}^L \hat{\sigma}^{z}_{j}+\frac{J }{2}\sum_{j=1}^{L-1}   ( \hat{\sigma}^{x}_{j} \hat{\sigma}^{x}_{j+1}+\hat{\sigma}^{y}_{j} \hat{\sigma}^{y}_{j+1}) ,\label{BH}
\end{eqnarray}
where $\hat{\sigma}_j^\alpha$ ($\alpha=x,y,z$) are the Pauli matrices with $j=0$ and $1,\cdots,L$, respectively, labeling the system spin and the spins in the chain; $\lambda $ denotes the longitudinal magnetic field exerted homogeneously on all the spins; $A(t)$ is the always-on periodic driving \cite{Cai2012, Jones2012} only on the system; $J$ and $g$ are, respectively, the coupling strengths between the nearest-neighbour spins of the chain and between the system and the first-site spin of the chain. $\hat{H}_\text{E}$ yields a phase transition at the critical point $|\lambda|=2J$ \cite{Katsura1962}. This type of system has been widely used to realize quantum state transfer, where the XX-coupling chain is used as a bridge \cite{Christandl2004, Zueco2009}, and to analyze decoherence caused by a spin bath \cite{Wang2012}. Diagonalizing $\hat{H}_\text{E}$ in the single-excitation subspace, we can obtain its eigenstate $|\varphi_k\rangle=\sum_{j=1}^{L}{e^{ikjx_0}\over \sqrt{L}}\hat{\sigma}^+_j|\{\downarrow_j\}\rangle$, which is a spin wave with wave vector $k$, and the eigenenergy $E_k=\lambda+2J\cos kx_0$ with $x_0$ being the spatial separation of the two neighbor sites. Here $ |\{\downarrow_j\}\rangle$ is the ferromagnetic state of the chain with all its spins pointing to the $-\hat{e}_z$ direction and $\hat{\sigma}_j^+=(\hat{\sigma}_j^x+i\hat{\sigma}_j^y)/2$. Obviously, the spin chain defines an environment with finite bandwidth $4J$.


We are interested in how the spin chain results in decoherence to the system spin and how it can be suppressed by periodic driving. Since the excitation number $\hat{\mathcal N}\equiv\sum_{j=0} ^{L} \hat{\sigma}^{+}_j\hat{\sigma}^{-}_j$ is conserved, the Hilbert space is divided into independent subspaces with definite $\mathcal{N}$. Consider that the spin chain is initially polarized in a ferromagnetic state and the system is in an up state $|\Psi(0)\rangle=|\phi\rangle \otimes |\{\downarrow_j\}\rangle$ with $|\phi\rangle=|\uparrow_0\rangle$, and its evolution can be expanded as $|\Psi(t)\rangle=e^{i{\frac{ L \lambda t}{2}}}\sum_{j=0}^Lc_j(t)\hat{\sigma}_j^+|\{\downarrow_j\}\rangle$, where $c_0(t)$ satisfies
\begin{eqnarray}
\dot{c}'_0(t)+i[\lambda+A(t)]c'_0(t)+\int_0^tf(t-\tau)c'_0(\tau)d\tau=0,\label{syspdec}
\end{eqnarray}with $c_0'(t)=c_0(t)e^{-{i\over 2}\int_0^t[\lambda+A(\tau)]d\tau}$ and $f(x)\equiv (g^2/L)\sum_k e^{-iE_kx}$ and $c_0(0)=1$. Denoting the excited-state probability of the system, $|c_0(t)|^2$ characterizes the environmental decoherence effect on the system. Equation \eqref{syspdec} provides us with the exact description to the decoherence of the system.

\section{Floquet quasi-energy spectrum}\label{results}
For a static system governed by $\hat{H}$, any time-evolved state can be expanded as
\begin{equation}
|\Psi(t)\rangle=\sum_nC_ne^{iE_nt}|\varphi_n\rangle        \label{static}
\end{equation}
where $C_n=\langle \varphi_n|\Psi(0)\rangle$, $E_n$ and $|\varphi_n\rangle$ determined by $\hat{H}|\varphi_n\rangle=E_n|\varphi_n\rangle$ are called as eigenenergies and stationary states, respectively.

A temporal periodic system governed by $\hat{H}(t)=\hat{H}(t+T)$ can be treated by Floquet theory \cite{Floquet1883}, which, as a powerful approach to map a non-equilibrium system under driving to a static one, can be seen as the application of Bloch theorem in the time domain. According to this theory, the periodic system has a complete set of basis $|u_{\alpha }(t)\rangle $ determined by
 \begin{equation}
\lbrack\hat{H}(t)-i\partial_t]|u_{\alpha}(t)\rangle =\epsilon_{\alpha }|u_{\alpha }(t)\rangle\label{flqe}
\end{equation}
such that any state can be expanded as
\begin{equation}|\Psi(t)\rangle=\sum_\alpha C_\alpha e^{-i \epsilon_\alpha t} |u_\alpha(t) \rangle \label{flqexp}
\end{equation}with $C_\alpha=\langle u_\alpha(0)|\Psi(0)\rangle$. The similar time-independence of $C_\alpha$ as $C_n$ in Eq. (\ref{static}) implies that $\epsilon _{\alpha}$ and $|u_{\alpha}(t)\rangle$ play the same roles in a periodic system as eigenenergies and stationary states do in static system. Such similarity leads us to call them quasienergies and quasi-stationary states, respectively. Carrying all the quasi-stationary-state characters, the quasienergy spectrum formed by all $\epsilon _{\alpha}$ is a key to study periodic system. Note that $\epsilon _{\alpha}$ is periodic with period ${2 \pi/ T}$ because $e^{i l\omega t} |u_\alpha(t)\rangle$ with $\omega={2\pi/T}$ is also the eigenstate of Eq. (\ref{flqe}) with eigenvalue $\epsilon_\alpha+l\omega$.

The Floquet operator acts on an extended Hilbert space named Sambe space, which is made up of the usual Hilbert space and an extra temporal space \cite{Shirley1965, Sambe1973}. To calculate the quasienergies, one first expands $|u_{\alpha}(t)\rangle$ in a complete set of basis of the temporal space, which is generally chosen as $\{e^{i k \omega t}|k\in Z\}$. We have $|u_\alpha(t)\rangle=\sum_{k}|\tilde{u}_{\alpha}(k)\rangle e^{ik\omega t}$, with which Eq. (\ref{flqe}) is recast into
\begin{equation}
\sum_{k\in Z}[\hat{\tilde{H}}_{l-k}+k\omega\delta_{l,k}]|\tilde{u}_\alpha(k)\rangle= \epsilon_\alpha|\tilde{u}_\alpha(l)\rangle,\label{flqt2}
\end{equation}with $\hat{\tilde{H}}_{l-k}\equiv\int_0^T \hat{H}(t){e^{-i(l-k)\omega t}\over T} dt$. Then expanding each $ \hat{\tilde{H}}_l$ in the complete basis of Hilbert subspace with ${\mathcal N}=1$, we get an infinite matrix equation. The quasienergies are obtained by truncating the basis of the temporal space to the rank such that the obtained magnitudes converge.

\section{Decoherence inhibition by periodic driving}\label{mechanism}
\subsection{The mechanism of the decoherence inhibition}
To reveal the mechanism of decoherence inhibition by the periodic driving, we consider explicitly that the energy splitting of the system is modulated as \cite{Wang2012}
\begin{equation}\label{MTF}
  A(t)=\left\{
    \begin{array}{cc}
      a_1, & n T<t \leq n T+ \tau  \\
      a_2, & ~~~n T+ \tau<t \leq (n+1)T
    \end{array}
  \right..
\end{equation}It is realizable by adding a time-dependent longitudinal magnetic field. Note that although only the driving periodic in this step function is considered, the mechanism revealed in the following is also applicable to other forms.
To Eq. (\ref{MTF}), we have
\begin{eqnarray}
\hat{\tilde{H}}_l&=&(\hat{ H}_\text{E}+\hat{H}_\text{I})\delta_{l,0}+(\omega_l/ 2) \hat{\sigma}_0^z,\\
\omega_l&=&{a_1(1-e^{-il\omega\tau})-a_2(e^{-i2\pi l}-e^{-i  l \omega \tau})\over2 i\pi l}.
\end{eqnarray}

\begin{figure}
  \includegraphics[width=0.9\columnwidth]{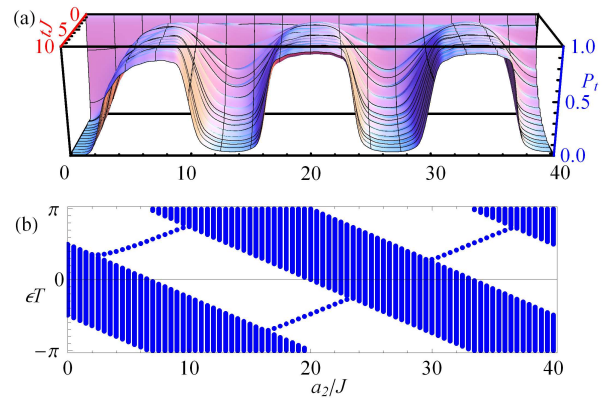}
  \caption{(Color online) (a) Evolution of the excited-state probability $P_t$ of the system spin in different driving amplitude $a_2$. (b) Floquet quasienergy spectrum of the whole system with the change of the driving amplitude $a_2$ in step $\delta a_2=0.5J$. The parameters $T=0.25\pi J^{-1}$, $a_1=0$, $\tau=0.1\pi J^{-1}$, $g=1.0J$, $\lambda=20.0J$, and $L=800$ are used. } \label{TTMFH}
\end{figure}
\begin{figure}\begin{center}
  \includegraphics[width=0.9\columnwidth]{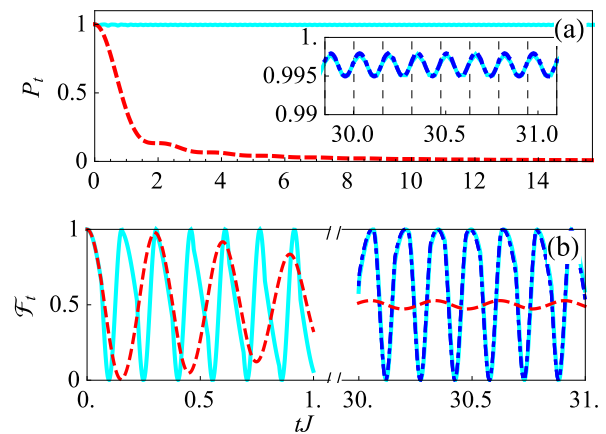}\end{center}
  \caption{(Color online) Evolution of $P_t$ for $|\phi\rangle=|\uparrow_0\rangle$ in (a) and $\mathcal{F}_t$ for $|\phi\rangle=(|\uparrow_0\rangle+|\downarrow_0\rangle)/\sqrt{2}$ in (b) when $a_2=36.0J$ with the FBS (cyan solid line) and $a_2=1.5J$ without the FBS (red dashed line) via numerically solving Eq. (\ref{syspdec}). The blue dotdashed lines show the results obtained via analytically evaluating the contribution of the FBS to the asymptotic state, which match with the numerical ones. The parameters are the same as Fig. \ref{TTMFH} except for $T=0.05\pi J^{-1}$ and $\tau=0.02\pi J^{-1}$.} \label{ddtal}
\end{figure}
We first study the asymmetric driving situation by choosing $a_1=0$. Figure \ref{TTMFH}(a) shows the time evolution of the excited-state probability $P_t=|c_0(t)|^2$ with the change of the driving amplitude $a_2$ via numerically solving Eq. \eqref{syspdec}. When the driving is switched off, i.e., $a_2=0$, $P_t$ decays monotonically to zero, which means a complete decoherence exerted by the spin chain to the system spin. When the driving is switched on, it is interesting to see that, dramatically different from the switch-off case, $P_t$ is stabilized repeatedly with the increase of $a_2$. To explain this, we plot in Fig. \ref{TTMFH}(b) the quasienergy spectrum obtained by solving Eq. \eqref{flqt2}. We can find that an FBS is possible to be formed within the bandgap with the increase of $a_2$. It is remarkable to see that the regimes where the decoherence is inhibited match well with the ones where the FBS is present. To understand the decoherence inhibition induced by the FBS, we, according to Eq. \eqref{flqexp}, rewrite
\begin{eqnarray}|\Psi(t)\rangle&=&e^{i{\frac{ L \lambda t}{2}}}[xe^{-i\epsilon_\text{FBS}t}|u_\text{FBS}(t)\rangle\nonumber\\
&&+\sum_{\alpha\in\text{Band}}y_\alpha e^{-i\epsilon_\alpha t}|u_\alpha(t)\rangle],\end{eqnarray} where $x=\langle u_{\text{FBS}}(0)|\Psi (0)\rangle $ and $y_{\alpha }=\langle u_{\alpha }(0)|\Psi (0)\rangle $. Then one can get that $P_t$ evolves asymptotically to $P_\infty\equiv x^2|\langle\Psi(0)|u_\text{FBS}(t)\rangle|^2$ with all the components in the quasi-energy band vanishing due to the out-of-phase interference contributed by the continuous phases (see Appendix \ref{app1}), as confirmed in Fig. \ref{ddtal}(a). In the absence of the FBS, although it is dramatically interrupted by the driving, $P_t$ decays to zero finally. Whenever the FBS is formed, $P_t$ would be stabilized to $P_\infty$, which is periodic with period $T$ [see the inset of Fig. \ref{ddtal}(a)]. It means that the presence of the FBS would cause $P_t$ to survive in the only component of the FBS and thus synchronize with the driving field \cite{Russomanno2012}. Figure \ref{ddtal}(b) plots the performance of the formed FBS in an arbitrary initial state $|\phi\rangle=(|\uparrow_0\rangle+|\downarrow_0\rangle)/\sqrt{2}$. We can see that the decay of the initial-state-fidelity $\mathcal{F}_t\equiv \langle \phi|\text{Tr}_\text{E}[|\Psi(t)\rangle\langle \Psi(t)|]|\phi\rangle$, to 50\% can be stabilized even as high as the ideal lossless case (i.e., between 0 and 1) with the formation of the FBS. Characterizing the quantum coherence between the two spin states, such stabilized oscillation means that the quantum coherence is preserved. We can check that $\mathcal{F}_t$ tends to $\mathcal{F}_\infty=\langle\phi|\rho|\phi\rangle$, where
\begin{eqnarray}
\rho&=&(1-{|x|^{2}\over2})|\downarrow _{0}\rangle \langle \downarrow
_{0}|+{|x|^{2}\over 2}\rho _{\text{FBS}}(t)\nonumber\\
&&+\{{x^{\ast }\over 2}\mu(t)\text{Tr}_{\text{E}}[|\{\downarrow
_{j}\}\rangle \langle u_{\text{FBS}}(t)|]+\text{h.c.}\}
\end{eqnarray} with $\rho _{\text{FBS}}(t)=\text{Tr}_{\text{E}}[|u_{\text{FBS}}(t)\rangle
\langle u_{\text{FBS}}(t)|]$ and $\mu(t)=e^{i\int_{0}^{t}\frac{\lambda
+A(t^{\prime })+2\epsilon _{\text{FBS}}}{2}dt^{\prime }}$ (see Appendix \ref{app2}). We plot this $\mathcal{F}_\infty$ with the blue dotdashed line in Fig. \ref{ddtal}(b), which matches with the asymptotical result from numerically solving Eq. (\ref{syspdec}).

\begin{figure}
  \includegraphics[width=0.95\columnwidth]{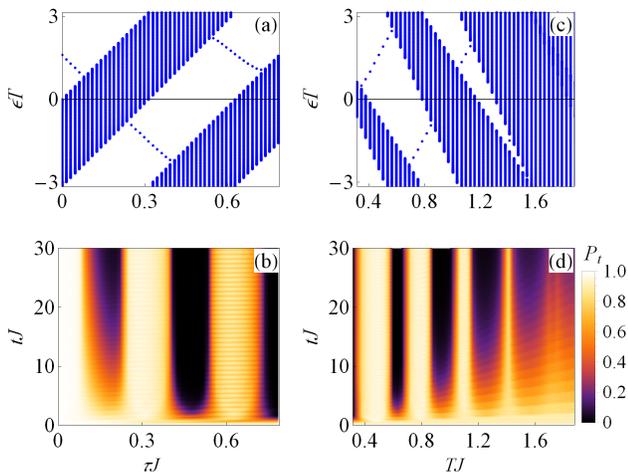}
  \caption{(Color online) Floquet quasienergy spectrum of the whole system in (a) and evolution of $P_t$ in (b) with the change of $\tau$ when $T=0.25\pi J^{-1}$. The increase step $\delta \tau=0.03J^{-1}$ is used in (a). Floquet quasienergy spectrum in (c) and time evolution of $P_t$ in (d) with the change of of $T$ when $\tau=0.1\pi J^{-1}$. The increase step $\delta T=0.09J^{-1}$ is used in (c). Other parameters are the same as Fig. \ref{TTMFH}. } \label{MT}
\end{figure}
The result reveals that we can manipulate the quasienergy spectrum forming the FBS to suppress decoherence. A prerequisite for forming the FBS is the existence of finite quasienergy gap in the spectrum. We plot in Fig. \ref{MT} the Floquet quasienergy spectrum and $P_t$ with the change of $\tau$ as well as $T$. We can see that, irrespective of which driving parameter is changed, the firm correspondence between the formation of the FBS and the decoherence inhibition can be established. The common character between Fig. \ref{TTMFH}(b) and Fig. \ref{MT}(a) is that the width of the formed bandgap is kept constant during the change of driving parameters, which is not true for Fig. \ref{MT}(c). This can be understood in the following way. Periodic in $2\pi/T$, the quasienergy has a full width $2\pi/T$. The energy band of the whole system is $4J$. Therefore, a bandgap with finite width $2\pi/T-4J$ can be present in the quasienergy spectrum only in the high-frequency (i.e. $2\pi/T>4J$) driving case. This can be tested by Fig. \ref{MT}(c) where the bandgap vanishes whenever $2\pi/T<4J$. It leads to the continuous energy band of the environment filling up the Floquet spectrum. Thus there is no room for forming the FBS here. Reflecting on $P_t$ in Fig. \ref{MT}(d), although it is greatly slowed, $P_t$ approaches zero eventually. Therefore, we conclude that the FBS can be present only in the high frequency driving case $2\pi/T>4J$, which supplies a necessary condition to stabilize decoherence. It is a very useful criterion on designing a driving scheme for decoherence control.

Our finding in the periodically driven system is an analog to the bound-state-induced decoherence suppression revealed in a static system \cite{Miyamoto2005, Tong2010, Zhang2013}. For a static two-level system \cite{Tong2010, Zhang2013} or a harmonic oscillator \cite{Lu2013} interacting with an environment, depending on the parameters in the spectral density, the total system may possess a stationary state named a bound state \cite{Tong2010} localized out of the continuous energy band of the environment. As a stationary state, the bound state contained as one superposition component in the initial state does not lose its quantum coherence during time evolution. Thus the system evolves exclusively to the time-invariant component of the bound state with other components in the continuous band vanishing due to their out-of-phase interference. This idea was used previously to suppress spontaneous emission of quantum emitters via introducing spatial periodic confinement to the radiation field in a photonic crystal setting \cite{Fujita2005, Englund2005, Jorgensen2011, Leistikow2011}. The spatial periodicity introduces a bandgap structure to the environmental energy spectrum such that an emitter-environment bound state is formed when the frequency of the emitter falls in the bandgap. Here we demonstrate that the parallel picture can be set up by introducing temporal periodicity to the system. The benefit of using the temporal periodic driving instead of the spatial periodic confinement is that its high controllability greatly relaxes the experimental difficulty in fabricating the spatial periodic confinement. Thus it is easier to realize in practice.

\begin{figure}[tbp]
  \includegraphics[width=0.95\columnwidth]{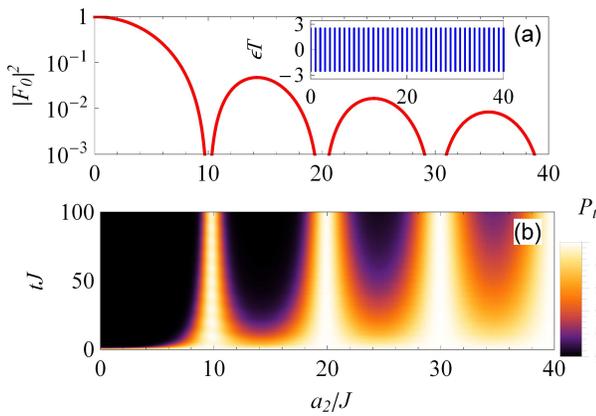}
  \caption{(Color online) The renomalization factor $|F_0|^2$ in (a) and the evolution of $P_t$ in (b) with the change of $a_2$ in the symmetric driving case (i.e. $a_1=-a_2$). The inset of (a) shows the Floquet quasienergy spectrum revealing the absence of the FBS. The parameters are $T=0.4\pi J^{-1}$ and $\tau=0.2\pi J^{-1}$ and the others are the same as Fig. \ref{TTMFH}.}\label{DD}
\end{figure}

\subsection{Comparisons with the previous methods}
There are several methods in the literature to explore the effects of periodic driving on quantum systems. For example, via neglecting the coupling between different temporal subspaces of the Floquet eigenequation \eqref{flqt2} in the high-frequency driving condition, it was shown that the periodic driving can induce the suppressed tunneling of a quantum particle, a phenomenon called coherent destruction of tunneling \cite{Grossmann1991,Grifoni1998,Luo2014}, and the decoupling between open system and its environment \cite{Zhou2009, Zhou2010}. It was also revealed that, via introducing the first-Markovian approximation to Eq. \eqref{syspdec}, the dynamics of the open system under periodic control can be characterized by an overlap integration of the noise spectrum and the spectrum of the control and thus one can craft the filter-transfer function of the control field to suppress decoherence \cite{KK2001}. It is called spectral filtering theory and has been generalized to give a unified description to a dynamical decoupling method \cite{KK2001, Uhrig2007, Sarma2008, Biercuk2011, Green2013}. In the following we compare our exact treatment with the above approximate methods.

First, our decoherence inhibition mechanism is more robust to the imperfect fluctuation of the driving parameters than the decoupling mechanism revealed in Ref. \cite{Zhou2009, Zhou2010}, where the decoupling is achieved only in certain single values of the driving parameters. To see this, we resort to the same approximate method as in Refs. \cite{Zhou2009, Zhou2010,Luo2014}. Expanding $|u_\alpha(t)\rangle$ in a new set of basis of the temporal space as $|u_\alpha(t)\rangle=\sum_k\hat{U}_te^{ik\omega t}|\tilde{u}_\alpha(k)\rangle\rangle$, where $\hat{U}_t=\exp[-(i/2)\int_0^t(A(t')-\bar{A})\hat{\sigma}_zdt']$ with $\bar{A}=(1/T)\int_0^TA(t)dt$ subtracted to guarantee the periodicity of $|u_\alpha(t)\rangle$, we can obtain a similar form as Eq. \eqref{flqt2} but
\begin{eqnarray}
\hat{\tilde{H}}_{l-k}&=&[{\lambda+\bar{A}\over 2}\hat{\sigma}_0^z+\hat{H}_\text{E}]\delta_{l,k}+ g(F_{l-k}\hat{\sigma}_1^+ \hat{\sigma}^-_0 +\text{h.c.}),~~~\\
F_{l-k}&=&\int_0^T{\exp\{-i\int_0^t[A(t')-\bar{A}]dt'\}e^{-i(l-k)\omega t}\over T}dt.
\end{eqnarray} Using the approximation in Refs. \cite{Zhou2009, Zhou2010}, we neglect the terms $F_{l-k}$ with $l\neq k$ and keep only $F_0$. It reduces to a spin system coupled to an environment with the coupling strength renormalized by a factor $F_{0}$. In Fig. \ref{DD}, we plot $|F_0|^2$ and $P_t$ with the change of $a_2$ in the symmetric driving situation, i.e. $a_1=-a_2$. It shows that although no FBS is formed, $F_0=0$ is achievable in certain values of driving parameters. As expected, it induces the decoherence inhibited [see Fig. \ref{DD}(b)]. However, the decoupling is sensitive to the driving parameters and any small deviation to the decoupling driving values would cause the asymptotic vanishing of $P_t$. Different from this, it is a wide parameter regime in our mechanism which makes decoherence inhibited (see Figs. \ref{TTMFH} and \ref{MT}), which is more stable to the parameter fluctuation in the practical experiments than the decoupling one.

\begin{figure}[tbp]
\includegraphics[width=.95\columnwidth]{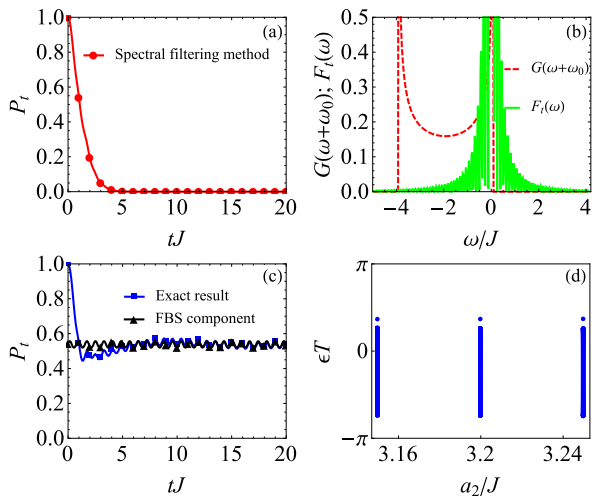}
\caption{(Color online) The comparison of $P_t$ calculated by the spectral filtering method
in (a) and our exact method in (c). (b): The noise spectrum $G(\omega+\omega_0)$ and the spectrum of the control $F_t(\omega)\equiv {|\varepsilon_t(\omega)|^2}/{Q(t)}$ used in the spectral filtering method to determine $P_t$. The contribution of the formed FBS to $P_t$ is also plotted in (c). The Floquet quasi-energy spectrum in (d) shows the existence of the FBS. $a_2=3.2J$ is further used in (a-c) and other parameters are same as Fig. \ref{TTMFH}.}
\label{CMP}
\end{figure}

Second, we emphasize that, our mechanism is substantially different from the spectral filtering theory \cite{KK2001, Uhrig2007, Sarma2008, Biercuk2011, Green2013}. That theory works only in the first-Markovian approximation, with which the convolution in the exact evolution equation (\ref{syspdec}) can be removed \cite{KK2001}, i.e.,
\begin{eqnarray}
\dot{\alpha}(t) \approx -\alpha (t)\int_{0}^{t}d\tau\varepsilon^* (t)\varepsilon
(\tau)f (t-\tau)e^{i\omega _{a}(t-\tau)},
\label{app}
\end{eqnarray}%
with $\alpha(t)=c'_0(t)e^{i\int_0^t[\lambda+A(\tau)]d\tau}$, $\varepsilon(t)=e^{i\int_0^t[A(\tau)-\bar{A}]d\tau}$, and $\omega_a=\lambda+\bar{A}$. Its solution can be obtained readily as
\begin{eqnarray}
|\alpha (t)|=|c_0(t)| = \exp [ -R(t)Q(t)/2] ,  \label{solu}
\end{eqnarray}%
where
\begin{eqnarray}R(t)&\equiv& 2\pi \int_{-\infty }^{+\infty
}G(\omega +\omega _{a}){\frac{|\varepsilon _{t}(\omega )|^{2}}{Q(t)}}d\omega,\\
Q(t)&=&\int_{0}^{t}d\tau |\varepsilon (\tau )|^{2}\end{eqnarray} with the environmental spectral density $G(\omega )$ relating to its
correlation function $f (t-\tau)$ as $f (t-\tau)=\int
G(\omega )e^{-i\omega (t-\tau)}d\omega $ and $\varepsilon _{t}(\omega )={%
\frac{1}{\sqrt{2\pi }}}\int_{0}^{t}\varepsilon (\tau)e^{i\omega
\tau}d\tau$. Thus it is only
under the first Markovian approximation that $|c_0 (t)|$ can be denoted
by such filtered spectrum form. To check the physics missed by this approximation, we plot in Fig. \ref{CMP} the comparison of our exact result with
the one obtained firmly from the spectral filtering theory. We can see from
Fig. \ref{CMP}(a) that the spectral filtering theory shows a complete
decoherence to zero because of a dramatic overlap between the environmental
spectrum and the control spectrum [see Fig. \ref{CMP}(b)]. However, our
exact result in Fig. \ref{CMP}(c) shows a stabilization on decoherence due to the existence of the FBS in the quasi-energy spectrum [see Fig. \ref{CMP}(d)]. It means that the
spectral filtering theory totally breaks down in describing the long-time
steady state behavior here. To give more evidence on the dominate role of
the formed FBS in the steady-state behavior, we plot in Fig. \ref{CMP}(c) the fidelity of the FBS in the
time-evolved state, which matches well with $P_t$ in the long-time limit.
Therefore, it confirms again that the formed FBS is the physical reason
for decoherence inhibition in long-time limit of our model. Thus the
decoherence cannot be simply described as an overlap between the noise
spectrum and the control field here and the spectral filtering theory is inapplicable to explain our result.

As a final remark, the mechanism revealed in our spin-bath model can also be readily extended to other excitation-number-conserving models, e.g. a two-level system in a coupled cavity array \cite{Zhou2009, Zhou2010} and a harmonic oscillator in a bosonic bath model \cite{Lu2013}.

\section{Conclusions} \label{conclusion}
We have studied the decoherence dynamics of a periodically driven spin-1/2 particle interacting with an XX coupled spin chain. It is found that the decoherence of the system can be inhibited by the periodic driving. We have revealed that the mechanism of such decoherence inhibition induced by the periodic driving is the formation of a FBS in the quasienergy spectrum. This can be seen as a close analog of the bound-state induced decoherence suppression in a photonic crystal system, but it relaxes greatly the experimental difficulties of a photonic crystal system in fabricating specific spatial periodicity to engineer a bound state. It opens a door to beat decoherence by tailoring temporal periodicity. Compared with the conventional schemes of decoherence control using periodic driving or pulses, our scheme is robust to the practical driving parameter fluctuation. Given the fact that periodic driving offers a high controllability to quantum system, our decoherence inhibition mechanism provides us with a promising and realistic way to practical decoherence control.

\section*{Acknowledgments}
This work is supported by the Fundamental Research Funds for the Central Universities, by the Specialized Research Fund for the Doctoral Program of Higher Education, by the Program for NCET, the National 973-program (Grant No. 2012CB922104 and No. 2014CB921403), by the NSF of China (Grants No. 11175072, No. 11174115, No. 11121403, No. 11325417, and No. 11474139), and by the National Research Foundation and Ministry of Education, Singapore (Grant No. WBS: R-710-000-008-271).

\appendix
\section{The contribution of the formed FBS to the long-time steady state}\label{app1}

For the initial state $|\Psi (0)\rangle =|\uparrow \rangle \otimes
|\{\downarrow _{1}\cdots \downarrow _{L}\}\rangle $, $|\Psi (t)\rangle $ can
also be expanded in the Floquet basis as
\begin{eqnarray}
|\Psi (t)\rangle&=&e^{i{\frac{ L \lambda t}{2}}}[xe^{-i\epsilon _{\text{%
FBS}} t}|u_{\text{FBS}}(t)\rangle\nonumber\\&&+\sum_{\alpha \in \text{B}}y_{\alpha }e^{-i\epsilon _{\alpha
}t}|u_{\alpha }(t)\rangle] ,  \label{QuasiEnerExp}
\end{eqnarray}%
where $|u_{\text{FBS}}(t)\rangle $ is the formed FBS with
quasienergy $\epsilon _{\text{FBS}}$, $|u_{\alpha }(t)\rangle $ are the
Floquet eigenstates in the continuous band with quasienergies $\epsilon
_{\alpha }$, $x=\langle u_{\text{FBS}}(0)|\Psi (0)\rangle $, and $%
y_{\alpha }=\langle u_{\alpha }(0)|\Psi (0)\rangle $. Then we can calculate the probability of the
system spin keeping in up state as
\begin{eqnarray}
P_{t} &=&|x|^{2}|\langle \Psi (0)|u_{\text{FBS}}(t)\rangle |^{2}  \notag \\
&&+\sum_{\alpha ,\beta \in \text{B}}y_{\alpha }^{\ast }y_{\beta
}e^{-i(\epsilon _{\beta }-\epsilon _{\alpha })t}\langle \Psi (0)|u_{\beta
}(t)\rangle \langle u_{\alpha }(t)|\Psi (0)\rangle  \notag \\
&&+\sum_{\alpha \in \text{B}}[xy_{\alpha }^{\ast }e^{-i(\epsilon _{\text{%
FBS}}-\epsilon _{\alpha })t}\langle \Psi (0)|u_{\text{FBS}}(t)\rangle
\langle u_{\alpha }(t)|\Psi (0)\rangle  \notag \\
&&+\text{c.c.}]
\end{eqnarray}%
Due to the out-of-phase interference contributed from $e^{-i(\epsilon
_{\beta }-\epsilon _{\alpha })t}$ with $\alpha \neq \beta $ and $%
e^{-i(\epsilon _{\text{FBS}}-\epsilon _{\alpha })t}$, $P_{t}$ tends to
\begin{eqnarray}
P_{\infty } &=&|x|^{2}|\langle \Psi (0)|u_{\text{FBS}}(\infty )\rangle
|^{2}+\sum_{\alpha \in \text{B}}|y_{\alpha }|^{2}|\langle \Psi
(0)|u_{\alpha }(\infty )\rangle |^{2}  \notag \\
&=&|x|^{2}|\langle \Psi (0)|u_{\text{FBS}}(\infty )\rangle |^{2}  \notag \\
&&+\sum_{\alpha \in \text{B}}|y_{\alpha }|^{4}|\langle u_{\alpha
}(0)|u_{\alpha }(\infty )\rangle |^{2}
\end{eqnarray}%
where the orthogonality of Floquet eigenstates has been used. Noticing the
fact that $\sum_{\alpha \in \text{B}}|y_{\alpha }|^{2}=\sum_{\alpha
=1}^{L}|y_{\alpha }|^{2}\sim 1$ (because we have $L$ Floquet eigenstates
forming the continuous quasienergy band), we can estimate that $|y_{\alpha
}|^{2}\sim 1/L$. In the thermodynamics limit $L\Rightarrow
\infty $, the last term tends to zero. Thus we have
\begin{eqnarray}
P_{\infty }&=&|x|^{2}|\langle \Psi (0)|u_{\text{FBS}}(\infty )\rangle |^{2}.
\end{eqnarray}

\begin{figure}[tbp]
\centering
\includegraphics[width=0.8\columnwidth]{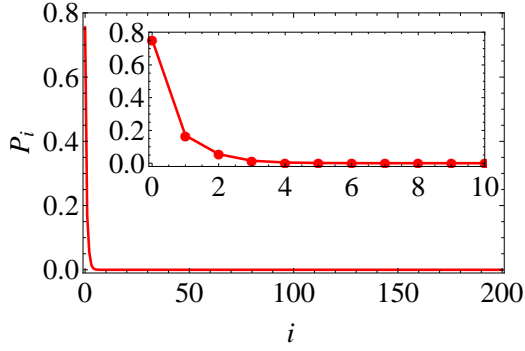}\newline
\caption{(Color online) The distribution of excited-state population of the formed Floquet
bound state at time $t=T/4$ over the spin sites. The parameters used are $%
T=0.25\protect\pi J^{-1}$, $\protect\tau=0.1\protect\pi J^{-1}$, $a_2=3.2J$,
and $a_1=0$.}
\label{FBSP}
\end{figure}
From the above analysis, we can see that the preserved excited-state
probability is determined by the weight of $|u_\text{FBS}(0)\rangle$ in
the initial state $|\Psi(0)\rangle$ and the excited-state probability of the system spin in $|u_%
\text{FBS}(\infty)\rangle$ itself. In Fig. \ref{FBSP}, we plot the distribution
of excited-state population of the formed FBS at time $t=T/4$
over the spin sites. We can see that its excited-state population is mainly
confined in the site of the system spin, which acts as an impurity in the
whole system.

\section{The effect of periodic driving on the initial superposition state}\label{app2}

For the general initial state $|\Psi(0)\rangle=|\phi%
\rangle\otimes |\{\downarrow_{j\neq 0}\}\rangle$ with $|\phi\rangle=\alpha
|\uparrow_0 \rangle + \beta |\downarrow_0 \rangle$ under $%
|\alpha|^2+|\beta|^2=1$, its evolved state $|\Psi(t)\rangle$ can be expanded
as
\begin{eqnarray}  \label{SupState}
|\Psi(t)\rangle&=&e^{i{\frac{ L \lambda t}{2}}} \Big( \alpha \sum_{j=0}
^{L}c_j(t) \hat{\sigma}_j^{+}|\{\downarrow_j\}\rangle  \notag \\
&&+ \beta e^{i\int_0 ^t \frac{\lambda+A(t^\prime)}{2} d
t^\prime}|\{\downarrow_j\}\rangle \Big),
\end{eqnarray}
where $c_0(t)$ satisfies Eq. (3) in the main text. The fidelity of the
system in its initial state $|\phi\rangle$ can be calculated as
\begin{eqnarray}  \label{ProbOfCohPres}
\mathcal{F}_t&=&\langle \phi|\text{Tr}_\text{E}[|\Psi(t)\rangle \langle
\Psi(t)|]|\phi\rangle  \notag \\
&=&\Big||\alpha|^2 c_0(t)e^{-i\int_0 ^t \frac{\lambda+A(t^\prime)}{2} d
t^\prime}+|\beta|^2\Big|^2  \notag \\
&&+|\alpha\beta|^2|[1-|c_0(t)|^2],
\end{eqnarray}
Since $|\phi\rangle$ is not an eigenstate of the system even in the
absence of the environmental influence, $\mathcal{F}_t$ is a temporally
oscillating function even in the long time limit. To qualitatively reflect
the performance of the periodic driving on suppressing decoherence, we use
the maximal value $\mathcal{F}_t$ to characterize it. This happens at a set
of times ${\tau_n}$ such that $c_0(\tau_n)e^{-i\int_0 ^{\tau_n} \frac{%
\lambda+A(t^\prime)}{2} d t^\prime}=|c_0(\tau_n)|$. Under this condition,
Eq. (\ref{ProbOfCohPres}) has the form
\begin{eqnarray}  \label{SimOfP}
\mathcal{F}_{\tau_n} &=&
1-|\alpha|^4[1-|c_0(\tau_n)|^2]-|\alpha\beta|^2[1-|c_0(\tau_n)|]^2  \notag \\
&\geq&|\beta|^2+|\alpha|^2|c_0(\tau_n)|^2.
\end{eqnarray}
When the FBS is absent, $|c_0(\infty)|=0$ and thus $\mathcal{F}%
_{\tau_n}=|\beta|^2 $. This corresponds to the complete decoherence (i.e,
the system spin decays totally to its low-energy spin down state). Whenever
the FBS is formed, a non-zero $|c_0(\infty)|$ would be achieved. Then we
could have $\mathcal{F}_{\tau_n}>|\beta|^2$ in the steady state. From this
analysis, we can see that the preserved probability for arbitrary initial
state is determined by the same long-time behavior of $|c_0(\infty)|$ as the
one for the spin up initial state. This proves well that our mechanism of
dissipation suppression can also be applied to the initial superposition
state.

More precisely, we can evaluate the contribution of the formed FBS to the steady state. $|\Psi (t)\rangle $ can also be expanded in the Floquet basis as
\begin{eqnarray}
&&|\Psi (t)\rangle =e^{i{\frac{L\lambda t}{2}}}\Big[\beta e^{i\int_{0}^{t}\frac{\lambda +A(t^{\prime })}{2}dt^{\prime
}}|\{\downarrow _{j}\}\rangle + \alpha (xe^{-i\epsilon
_{\text{FBS}}t}\notag \\
&&~~~\times |u_{\text{FBS}}(t)\rangle+\sum_{\gamma \in \text{B}%
}y_{\gamma }e^{-i\epsilon _{\gamma }t}|u_{\gamma }(t)\rangle ) \Big].
\end{eqnarray}%
Due to the out-of-phase interference, the reduced density matrix tends to
\begin{eqnarray}
\rho (\infty ) &=&\text{Tr}_{\text{E}}[|\Psi (\infty )\rangle \langle \Psi
(\infty)|] =|\beta |^{2}|\downarrow _{0}\rangle \langle \downarrow _{0}|  \notag
\\
&&+|\alpha |^{2}\{|x|^{2}\rho _{\text{FBS}}(t)+\sum_{\gamma }|y_{\gamma
}|^{2}\text{Tr}_{\text{E}}[|u_{\gamma }(t)\rangle \langle u_{\gamma }(t)|]\}
\notag \\
&&+\{\beta \alpha ^{\ast }x^{\ast }\mu(t)\text{Tr}_{\text{E}}[|\{\downarrow
_{j}\}\rangle \langle u_{\text{FBS}}(t)|]+\text{h.c.}\} ,
\end{eqnarray}%
where $\rho _{\text{FBS}}(t)=\text{Tr}_{\text{E}}[|u_{\text{FBS}}(t)\rangle
\langle u_{\text{FBS}}(t)|]$ and $\mu(t)=e^{i\int_{0}^{t}\frac{\lambda
+A(t^{\prime })+2\epsilon _{\text{FBS}}}{2}dt^{\prime }}$. Noticing the fact
that Tr$_{\text{E}}[|u_{\gamma }(t)\rangle \langle u_{\gamma }(t)|]$ is
dominated by $|\downarrow _{0}\rangle \langle \downarrow _{0}|$ and $%
\sum_{\gamma }|y_{\gamma }|^{2}+|x|^{2}=1$, we have $\sum_{\gamma
}|y_{\gamma }|^{2}$Tr$_{\text{E}}[|u_{\gamma }(t)\rangle \langle u_{\gamma
}(t)|]\approx (1-|x|^{2})|\downarrow _{0}\rangle \langle \downarrow _{0}|$.
Thus the asymptotic state of the system spin is
\begin{eqnarray}
&&\rho (\infty ) =(1-|\alpha |^{2}|x|^{2})|\downarrow _{0}\rangle \langle
\downarrow _{0}|+|\alpha |^{2}|x|^{2}\rho _{\text{FBS}}(t)  \notag \\
&&+\{\beta \alpha ^{\ast }x^{\ast }\mu(t)\text{Tr}_{\text{E}}[|\{\downarrow
_{j}\}\rangle \langle u_{\text{FBS}}(t)|]+\text{h.c.}\}.
\end{eqnarray}
Then the analytical form of the fidelity in the long-time limit can be
calculated by $\mathcal{F}_\infty=\langle\phi|\rho(\infty)|\phi\rangle$. It
gives the contribution of the formed FBS to the asymptotical state and can be used to check the validity of our FBS theory in explaining the dynamics of the system spin.


\end{document}